\documentclass[twocolumn,aps]{revtex4}
\usepackage{graphicx}
\draft

\newcommand{\nl}{\nonumber \\}

\newcommand{\be}{\begin{equation}}
\newcommand{\ee}{\end{equation}}
\newcommand{\bea}{\begin{eqnarray}}
\newcommand{\eea}{\end{eqnarray}}

\newcommand{\Eq}[1]{Eq.\,(\ref{#1})}

\newcommand{\la}{\langle}
\newcommand{\ra}{\rangle}
\newcommand{\dg}{\dagger}

\newcommand{\bfsigma}{\mbox{\boldmath $\sigma$}}

\newcommand{\mbB}{\mbox{\boldmath $B$}}
\newcommand{\mbe}{\mbox{\boldmath $e$}}

\newcommand{\mb}{\mbox}

\begin{document}
\draft

\title{ Ultrafast geometric manipulation of electron spin
and detection of the geometric phase via Faraday rotation spectroscopy }

\author{Xin-Qi Li, Cheng-Yong Hu, Li-Xiang Cen, and Hou-Zhi Zheng}
\address{National Laboratory for Superlattices and Microstructures,
         Institute of Semiconductors,
         Chinese Academy of Sciences, P.~O.~Box 912, Beijing 100083, China }
\author{YiJing Yan}
\address{Department of Chemistry, Hong Kong University of Science and Technology,
         Kowloon, Hong Kong }

\date{\today}

\begin{abstract}
Time-resolved Faraday rotation spectroscopy is currently
exploited as a powerful technique to probe spin dynamics in semiconductors.
We propose here an all-optical approach to geometrically manipulate
electron spin and to detect the geometric phase
by this type of extremely sensitive experiment.
The global nature of the geometric phase can make the quantum
manipulation more stable, which may find interesting application in
quantum devices.
\end{abstract}
\maketitle



Faraday rotation (FR) describes the phenomenon of the polarization of
a linearly polarized light, after transmitting through a magnetized medium,
being rotated around the light propagation or wave-vector direction.
FR results from the refraction index difference
between the left- and right-circularly polarized light,
and the amount of FR is proportional to
the sample magnetization along the light propagation direction.
Therefore, FR can be exploited as a non-invasive and sensitive means to measure material
magnetization.
In addition, the extreme sensitivity of the technique allows to develop
an efficient method (i.e. the time-resolved pump-probe approach)
to monitor and measure the spin dynamics of
photoexcited electrons, holes, and excitons in semiconductors.\cite{Cro9597,Gup01}

\begin{figure}\label{Fig1}
\begin{center}
\centerline{\includegraphics [bb= 26 556 574 766, clip,scale=0.4]{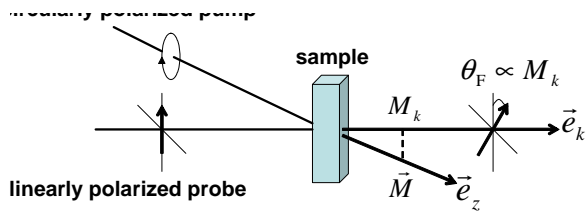}}
\caption{Schematic diagram for the principle of Faraday rotation spectroscopy. }
\end{center}
\end{figure}

In this paper, we elaborate a scheme of ultrafast geometric manipulation of
electron spin in semiconductor by an all-optical approach, and detection
of the geometric phase via FR spectroscopy.
Geometric phases, either the adiabatic Berry phase\cite{Ber84} or
the non-adiabatic Aharonov-Anandan (AA) phase,\cite{Aha87}
are of fundamental interest in physics.
Their existences had been experimentally demonstrated in many physical systems,
such as molecules,\cite{Del86} light polarization in fiber,\cite{Chi86}
NMR,\cite{Tyc87,Sut87,Sut88} and neutron polarization.\cite{Bit87}
In particular, also based on the FR principle,
a scheme to determine the geometric phase (with a unique value $\pi$)
was proposed, \cite{Zak91} where the geometric phase is associated with
the light polarization vector, and the FR turns out to be an optical analogue
of a spin-1/2 particle in a magnetic field.
More recently, in the context of geometric
quantum computation,\cite{Zan99}
both the conditional Berry and AA phases were measured
in liquid NMR system,\cite{Jon00,Du02}
and were proposed to construct quantum logic gate.
This interest is rooted in the fact that
the quantum manipulation by geometric means seems to be
fault-tolerant to certain types of computational errors.
We also notice that, with the advent of spintronics,
there exists rapidly growing interest in coherent control of spins in semiconductors.
The spin degree of freedom with its coherent control in solid states is widely
believed to have sound potential in quantum device applications.
We thus anticipate that the study of the geometric manipulation of electron spin
and the detection of the geometric phase via the extremely sensitive FR spectroscopy
would be relevant to this remarkable trend.

   Figure 1 depicts the time-resolved pump-probe FR spectroscopic setup.\cite{Cro9597,Gup01}
Here, the pump laser pulse, which is circularly polarized,
is applied first to create electron-hole pairs in
the semiconductor sample. As a result of the selection rule imposed by
angular momentum conservation, only the electron-hole pairs
consistent with the photon polarization can be excited.
Owing to the distinct relaxation timescales (the hole spin relaxation
is typically much faster),
the hole- and electron-spin relaxations are usually addressed separately.
As a good approximation, in this work we only consider the effect of electron
spins (in the conduction band).
Conventionally, the photoexcited carrier (conduction electron) spins are subjected to
the action of magnetic field, and the time-resolved FR signal reveals the motion of spins.
In this work, we are interested in the spin motion involving a geometric phase
that in turn generates observable effect in the FR signal.
Viewing the typical electron spin relaxation time of nanoseconds,\cite{Cro9597,Gup01}
magnetic field pulses shorter than this timescale are needed in order to
implement the desired cyclic quantum evolution of spin state.
Unfortunately, such short magnetic field pulses are beyond the scope of current technology.
Very recently, an all-optical approach
to the ultrafast manipulation of electron spin
has been demonstrated by making use
of the optical ac Stark effect that produces ultra-short
effective magnetic field pulses.\cite{Gup01}
In what follows, we first elaborate the principle of geometrically manipulating
the electron spin and using FR to detect the
geometric phase, then discuss the possible experimental implementation
by an all-optical approach.

For the sake of generality,
assume a mixed initial state of the carrier spins prepared by the pump laser
in the conduction band (the hole spins in the valence band can be ignored
because of their much shorter relaxation time)
\bea\label{rhoi}
\rho_i= w_0|0\ra\la 0| + w_1|1\ra\la 1|.
\eea
Here, $|0\ra$ and $|1\ra$ denote, respectively, the spin-down
and spin-up states of the excited electron,
with respect to the pump laser direction along the $z$-axis (see Fig.\ 1).
Correspondingly, we define the $x$-axis in the plane determined by
the pump and probe laser propagating directions,
and the $y$-axis perpendicular to this plane.
The time evolution of density operator is governed by the Liouville equation,
that results in a formal solution
\bea\label{rhot}
\rho(t)=U(t)\rho_i U^{\dg}(t),
\eea
where the operator $U(t)$ describes the quantum evolution starting from the initial
(mixed) state $\rho_i$.

\begin{figure}[b]\label{Fig2}
\begin{center}
\centerline{\includegraphics [scale=0.3]{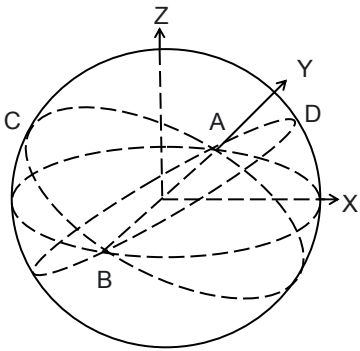}} \caption{
Schematic diagram for geometric rotation of state vector around
the magnetic field. Since the state is always perpendicular to the
field, there is no dynamical phase accumulation during the
evolution. }
\end{center}
\end{figure}

To evolve the electron spin state,
assume a magnetic field applied in the $x$-$z$ plane with, correspondingly,
two projective components $B_x$ and $B_z$.
For clarity, we first consider the non-adiabatic geometric evolution of the eigenstates
of $\sigma_y$, $\sigma_y|\pm\ra=\pm|\pm\ra$, where
$\sigma_y=\bfsigma\cdot\mbe_y$ is the Pauli spin operator defined along the
$\mbe_y$ direction ($y$-axis).
Specifically, the quantum cyclic evolution is accomplished by
the following magnetic field pulses:
(i) Switching on a $\pi$-pulse of magnetic field with components $B_z$ and $B_x$,
the state $|+\ra$ rotates around the magnetic field,
from $|+\ra$ in the $\mbe_y$ direction to $|-\ra$ in the $-\mbe_y$ direction
along curve $ACB$ on the Bloch sphere, see Fig.\ 2.
(ii) Suddenly changing the magnetic field to another direction with components
$B_z$ and $-B_x$, subjecting to another $\pi$-pulse action
the state $|-\ra$ rotates back to $|+\ra$ along curve $BDA$ on the Bloch sphere.
From the AA phase theory,\cite{Aha87}  after the above cyclic evolution, the state
$|+\ra$ will acquire a geometric AA phase $e^{i\gamma}$, with
$\gamma=4\mb{arctan}(B_x/B_z)$.
Note that during the above operation, the state vector is always being perpendicular
to the magnetic field, thus no dynamic phase is accumulated in the evolution.
This idea was first promoted by Suter {\it et al} in their seminal experiment of
demonstrating the AA phase.\cite{Sut88}
Similarly, the state $|-\ra$ will acquire an AA phase $e^{-i\gamma}$ at the same time.

In analogy with the Mach-Zender interferometer where the photon wavefunction
is splitted at the first beamsplitter, we consider the consequence of the above rotation
on the eigenstates of $\sigma_z$:
$|0\ra=\frac{-i}{\sqrt{2}}(|+\ra-|-\ra)$,
and $|1\ra=\frac{1}{\sqrt{2}}(|+\ra+|-\ra)$.
Straightforwardly, the quantum state evolution is described by
\bea\label{WF1}
|0\ra & \rightarrow &  \mb{cos}\gamma|0\ra+\mb{sin}\gamma|1\ra ,  \nl
|1\ra & \rightarrow &  \mb{cos}\gamma|1\ra-\mb{sin}\gamma|0\ra .
\eea
We see here that the geometric AA phase plays a role of rotating
$|0\ra$ and $|1\ra$ into their superposition states.
As a consequence, geometric rotation of the electron spin is realized.
As emphasized in the geometric quantum computation,\cite{Zan99}
the major advantage of geometric manipulation is its ability
to be fault-tolerant to certain types of errors.
Note also that the possible value of $\gamma$ ranges from $0$ to $2\pi$
by controlling the magnetic field components $B_x$ and $B_z$,
implying an arbitrary geometric rotation of the spin state.

Below we show that the geometric phase can be detected by FR.
In the representation of basis states $\{|1\ra,|0\ra \}$,
the evolution described by \Eq{WF1} can be equivalently reexpressed
in terms of the evolution operator (matrix) as
\bea\label{UT}
U(T)=\left(  \begin{array}{cc}
               \mb{cos}\gamma   &  \mb{sin}\gamma  \\
               -\mb{sin}\gamma  &  \mb{cos}\gamma
             \end{array}
     \right) ,
\eea
where $T$ is the entire geometric rotation time.
Substituting \Eq{UT} into (\ref{rhot}), the final state of the conduction
electron spins after the above geometric operation reads
\bea\label{rhof}
\rho_f &=& U(T)\rho_iU^{\dg}(T)   \nl
       &=& \left(  \begin{array}{cc}
  w_1\mb{cos}^2\gamma+w_0\mb{sin}^2\gamma & (w_0-w_1)\mb{cos}\gamma\mb{sin}\gamma \\
  (w_0-w_1)\mb{cos}\gamma\mb{sin}\gamma & w_1\mb{sin}^2\gamma+w_0\mb{cos}^2\gamma
             \end{array}
     \right) .
\eea
With this result,
any physical quantities can be carried out by statistically averaging
the corresponding variables over the system density matrix.
In the context of FR experiment, the relevant quantity
is the projection of sample magnetization in the propagating direction ($\mbe_k$) of the probe light.
The corresponding variable operator is
$\sigma_k=\bfsigma\cdot\mbe_k
 =\cos\alpha\sigma_z-\sin\alpha\sigma_x$,
where $\alpha$ is the angle between the pump- and probe-laser propagating directions.
Straightforwardly, the component along $\mbe_k$ of the sample magnetization is
calculated as
\bea\label{Mk}
M_k\propto \mb{Tr}(\sigma_k\rho_f)=(w_1-w_0)\mb{cos}(\alpha-2\gamma).
\eea
Two valuable observations on \Eq{Mk} are in order:
(i) The geometric phase $\gamma$ has clear observable effect in $M_k$,
thus in FR experiment. The dependence of FR angle $\theta_F$ ($\propto M_k$)
on the geometric phase $\gamma$ is in terms of the typical behavior of quantum
interferences. In the limit of $\gamma=0$, \Eq{Mk} reduces to the result
of FR experiment that probes the initially excited carrier spins
in the semiconductor sample.
(ii) The output intensity is an incoherently weighted average of the pure state
interference profiles. It is easy to check that for pure state $|1\ra$ and
$|0\ra$, the interference profiles are, respectively, $\pm\mb{cos}(\alpha-2\gamma)$.
The weight factors (i.e.,$w_1$ and $w_0$) are given in the initial state
$\rho_i$. This structure is consistent with the recent work on
geometric phases for mixed states in interferometry. \cite{Sjo00}

The interferometry described above is also applicable
to detect the adiabatic Berry phase, \cite{Ber84}
provided the magnetic field can adiabatically complete a closed path
in parameter space within the spin relaxation time.
This condition can be satisfied in some spin systems,
such as the liquid NMR or doped spins in solid-states.
Particularly, if the initial state is an eigenstate of the system Hamiltonian,
the adiabaticity will ensure that
the state in the subsequent time will remain in the same eigenstate of the
instantaneous Hamiltonian $H(\mbB)$ ($\mbB$ is the time-dependent magnetic field).
As $\mbB$ traces a closed loop in the parameter space, a geometric phase is acquired.
For the geometric-phase-based interference study,
the initial state $|0\ra$ (or $|1\ra$) is a
linear superposition of the eigenstates $|\pm\ra$ of $H(\mbB)$,
if the magnetic field is initially along the $y$-direction.
As a result of adiabatically dragging the Hamiltonian along a closed loop,
a phase difference between $|+\ra$ and $|-\ra$ is caused.
Noticeably, the phases acquired in this way have both the geometric and dynamic
contributions.
To detect merely the geometric Berry phase via an interferometry, it is necessary
to eliminate the dynamic phase.
An applicable approach is to use the refocusing technique known as spin-echo
developed in NMR experiment, \cite{Ern87}
which was also employed recently to eliminate the
dynamic phases in geometric quantum computation.\cite{Jon00}
The basic idea is to apply the cyclic evolution twice, with
the second cyclic evolution retracing the first one but following the reversed path,
and with the second application
surrounded by a pair of fast $\pi$-transformations that swap the
states $|+\ra$ and $|-\ra$.
The net effect of this compound operation is to
cancel the dynamic phases, and maintain the geometric ones.
Supposing that the net geometric phases acquired by $|\pm\ra$ are
$e^{\pm i\gamma_B}$, the interference pattern is analogous to the
previous non-adiabatic one, only with $\gamma$ in \Eq{Mk} being replaced
by $\gamma_B$.


Now let us return to the possible experimental implementation
of the ultrafast geometric manipulation
of electron spin and the detection of the non-adiabatic AA phase,
which is the major concern of the present study.
As we have mentioned earlier, a necessary condition for
realizing the geometric manipulation
is the cyclic evolution being much faster than the spin relaxation.
Viewing that the conduction-electron spin relaxation time is about tens of
nanoseconds, we thus need ultra-short magnetic field pulses.
We show below how this goal can be achieved
by virtue of the recent experiment on the ultrafast manipulation of electron spin
by a novel all-optical approach.\cite{Gup01}

\begin{figure}[b]\label{Fig3}
\begin{center}
\centerline{\includegraphics [scale=0.4]{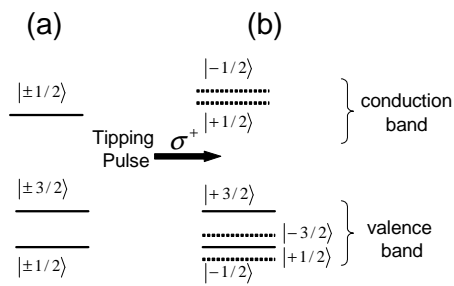}} \caption{
ac Stark shift induced by an off-resonance below-band-gap laser pulse.
Depicted are, respectively, the original un-perturbed energy levels (a)
and the shifted ones (b). }
\end{center}
\end{figure}

The basic idea is to make use of an off-resonance ultrafast laser pulse to induce ac Stark shifts,
which are in turn equivalent to the result of an effective magnetic field.
To illustrate this, consider the energy level diagram of a semiconductor
quantum well shown in Fig.\ 3(a).
In the absence of magnetic field, the lowest conduction-band (CB) level is two-fold degenerate,
denoted by spin states $|\pm 1/2\ra$; and
the valence-band (VB) states are denoted, respectively, by $|\pm 3/2\ra$ and $|\pm 1/2\ra$.
Switching on a below-band-gap laser pulse (tipping pulse) with, for example,
$\sigma^+$-polarization, the laser will virtually couple the state pairs between
VB state $|-3/2\ra$ and CB state $|-1/2\ra$, as well as CB state $|+/2\ra$ and  VB state $|-1/2\ra$.
Other possible couplings between laser and electron-hole pairs are forbidden
from the standpoint of angular momentum conservation.
It is well known that this kind of off-resonance coupling will cause the ac Stark shifts
of the relevant energy levels.
In particular, the energy shifts can be estimated from
the second-order perturbation theory as follows:
$\Delta E^c_{-1/2}=|V_{-3/2,-1/2}|^2/\Delta_1$ for CB state $|-1/2\ra$,
$\Delta E^v_{-3/2}=|V_{-3/2,-1/2}|^2/(-\Delta_1)$ for VB state $|-3/2\ra$,
$\Delta E^c_{1/2}=|V_{-1/2,+1/2}|^2/\Delta_2$ for CB state $|+1/2\ra$, and
$\Delta E^v_{-1/2}=|V_{-1/2,+1/2}|^2/(-\Delta_2)$ for VB state $|-1/2\ra$.
Here, $\Delta_1$ and $\Delta_2$ are the detunings of the photon energy with
the two pair states; $V_{-3/2,-1/2}$ and $V_{-1/2,+1/2}$ are coupling matrix elements
of laser with the state pairs.
Plotted in Fig.\ 3(b) are the shifted levels, with respect to the original
un-perturbed ones in Fig.\ 3(a).
We notice here that the level repulsive effect is obvious.

As experimental studies have indicated that the hole spins are either
pinned along the quantum well growth direction or dephase rapidly ,\cite{Cro9597,Gup01}
the FR dominantly measures the net effect of the CB electron spins.
As a consequence, the ac Stark shifts of CB states $|-1/2\ra$ and $|+1/2\ra$
can be equivalently described as
the effect of an effective magnetic field $B_{\mb{eff}}$,
along the tipping-pulse direction.
Denoting the CB level splitting by $\delta_{\mb{cb}}$,
the effective magnetic field $B_{\mb{eff}}=\delta_{\mb{cb}}/(g_c\mu_B)$,
where $g_c$ is the Lande-$g$ factor, and $\mu_B$ is the Bohr magneton.
As a rough estimate, corresponding to a CB level splitting of $\delta_{\mb{cb}}\sim 1$ meV,
the effective magnetic field $B_{\mb{eff}}$ can be as high as $\sim 20$ T.
In addition, the persistence time of this effective magnetic field is identical
to the laser-pulse duration time, that can be as short as femtoseconds.
Therefore, the resulting effective magnetic field can suffice
the geometric operation (rotation) of the electron spin under study.
To obtain the AA phase resulting interference pattern, the FR experiment
can be arranged as follows:
(i) After optical excitation using an ultrafast circularly polarized laser pulse,
the initial spin state is prepared along the $z$-direction shown in Fig.\ 1.
In the absence of magnetic field,
the FR  angle of the linearly polarized probe laser keeps unchanged.
(ii) Switch on two tipping pulses in succession such that
the effective magnetic field induced by the ac Stark effect rotates the
electron spin in the manner as shown in Fig.\ 2.
Note that the direction of the effective magnetic field is given by
the tipping pulse propagating direction, and its magnitude depends on the
detunings and the coupling strengths between the tipping laser and electron-hole pairs.
After the action of the tipping pulses, the FR angle will gain a change as show by \Eq{Mk}.
(iii) Repeat procedures (i) and (ii) by making changes of the tipping pulse directions.
A series of values of the AA phase $\gamma$ can be measured,
and accordingly the interference pattern resulting from the geometric AA phase
can be obtained.


In summary, we have elaborated an experimentally accessible scheme
to geometrically manipulate electron spin, and to detect
the non-adiabatic geometric phase via FR spectroscopy.
This study might be of fundamental interest and relevant to the notion of spintronics.
Viewing the particular significance of ultrafast manipulation of
electron spin, the proposed study can further
confirm the novel approach developed in ref.\ \onlinecite{Gup01}.
Moreover, the proposed geometric scheme may further stabilize the
quantum operation, i.e., being of fault-tolerance to some operational errors.
Finally,
the proposed interference scheme based on the FR setup can also be applied
to detect geometric phases in other spin systems, including the
adiabatic Berry phase as long as the spin relaxation time is
longer than the experimentally accessible magnetic field duration time.

\vspace{3ex}
{\it Acknowledgments.}
  Support from the Major State Basic Research Project
No.\ G001CB3095 of China, the Special Fund for``100 Person Project"
from Chinese Academy of Sciences,
and the Research Grants Council of the Hong Kong Government
are gratefully acknowledged.


\end{document}